\documentclass[twocolumn, table]{aastex631}
\usepackage{color}
\usepackage{graphicx}
\usepackage{amsmath}
\usepackage{subfigure}

\newcommand{\rmsub}[1]{\ensuremath{_{\mathrm{#1}}}}

\newcommand{\yr}{\textrm{yr}} 
\newcommand{\ps}{\ensuremath{\mathrm{s^{-1}}}} 
\newcommand{\km}{\textrm{km}} 
\newcommand{\pc}{\textrm{pc}} 

\shorttitle{Timing analysis of the pulsars in the globular cluster M53}
\shortauthors{Lian, et al}

\begin{document}

\title{Discovery and Timing analysis of new pulsars in globular cluster NGC 5024: new observations from FAST}

\author{Yujie Lian}
\affiliation{Institute for Frontiers in Astronomy and Astrophysics, Beijing Normal University, Beijing 102206, China;}
\affiliation{Department of Astronomy, Beijing Normal University, Beijing 100875, China;}

\author{Zhichen Pan$^{\ast}$}
\affiliation{National Astronomical Observatories, Chinese Academy of Sciences, 20A Datun Road, Chaoyang District, Beijing, 100101, China;}
\affiliation{CAS Key Laboratory of FAST, National Astronomical Observatories, Chinese Academy of Sciences, Beijing 100101, China;}
\affiliation{College of Astronomy and Space Sciences, University of Chinese Academy of Sciences, Beijing 100049, China;}

\author{Haiyan Zhang$^{\dag}$}
\affiliation{National Astronomical Observatories, Chinese Academy of Sciences, 20A Datun Road, Chaoyang District, Beijing, 100101, China;}
\affiliation{CAS Key Laboratory of FAST, National Astronomical Observatories, Chinese Academy of Sciences, Beijing 100101, China;}
\affiliation{College of Astronomy and Space Sciences, University of Chinese Academy of Sciences, Beijing 100049, China;}

\author{P. C. C. Freire}
\affiliation{Max-Planck-Institut fur Radioastronomie, Auf dem Hugel 69, D-53121 Bonn, Germany}

\author{Shuo Cao$^{\ddag}$} 
\affiliation{Institute for Frontiers in Astronomy and Astrophysics, Beijing Normal University, Beijing 102206, China;}
\affiliation{Department of Astronomy, Beijing Normal University, Beijing 100875, China;} 

\author{Lei Qian}
\affiliation{National Astronomical Observatories, Chinese Academy of Sciences, 20A Datun Road, Chaoyang District, Beijing, 100101, China;}
\affiliation{CAS Key Laboratory of FAST, National Astronomical Observatories, Chinese Academy of Sciences, Beijing 100101, China;}
\affiliation{College of Astronomy and Space Sciences, University of Chinese Academy of Sciences, Beijing 100049, China;}

\footnote{$\ast$ Equal contribution; panzc@bao.ac.cn}
\footnote{$\dag$ Equal contribution; hyzhang@bao.ac.cn}
\footnote{$\ddag$ caoshuo@bnu.edu.cn}

\begin{abstract}

NGC 5024 (M53) is the most distant globular cluster (GC) with known pulsars. In this study, we report the discovery of a new binary millisecond pulsar PSR J1312+1810E (M53E) and present the new timing solutions for M53B to M53E, based on 22 observations from the Five-hundred-meter Aperture Spherical radio Telescope (FAST). These discoveries and timing work benefit from FAST's high sensitivity. We find that M53C is the only isolated millisecond pulsar known in this distant globular cluster, with a spin period of 12.53 ms and spin period derivative of $5.26\times 10^{-20} \, \rm s \; s^{-1}$.  Our results reveal the orbital periods of 47.7, 5.8, and 2.4 days for M53B, D, and E, respectively. The companions, with a mass of 0.25, 0.27, and 0.18 ${\rm M}_\odot$, respectively, are likely to be white dwarf stars; if they are extended objects, they don't eclipse the pulsars. We find no X-ray counterparts for these millisecond pulsars in archival $Chandra$ images in the band of 0.3–8 keV. The characteristics of this pulsar population are similar to the population of millisecond pulsars in the Galactic disk, as expected from the low stellar density of M53.

\end{abstract}


\section{Introduction}
\label{sec:introduction}

Globular clusters have such extraordinarily dense stellar environments that collisions and interactions between stars are very likely over their lifetimes. For this reason, they form unusually large numbers of low-mass X-ray binaries (LMXBs) and the latter's evolutionary products, millisecond pulsars (MSPs) per unit stellar mass \citep{Clark1975,Katz1975}. The LMXBs form when a member of a binary is exchanged by a lone neutron star in an {\em exchange interaction} \citep{Sigurdsson1993}.

For this reason, GCs are excellent places to search for MSPs. Following the discovery of first GC pulsar PSR B1821$-$24A in the GC M28 \citep{Lyne1987},  288 pulsars in 38 GCs have been reported until 2023 May\footnote{\url{http://www.naic.edu/~pfreire/GCpsr.html}}. This population is very different from the general pulsar population: The ATNF pulsar catalog has reported 533 MSPs with a spin period shorter than 20 ms \citep{Manchester2005}\footnote{\url{https://www.atnf.csiro.au/research/pulsar/psrcat/}}, which are less than 16\% of the total pulsar population. Meanwhile, in GCs this number is 259, about 90 \% of the total. In GCs, $\sim$ 56\% of pulsars are in binary systems. 
Some of these pulsars can be very unusual, including the fastest spinning pulsar J1748$-$2446ad (spin period 1.39595 ms; \citealt{Hessels2006})  and the MSPs with eccentric orbits and massive companions (e.g., PSR~J0514$-$4002A, which has an orbital eccentricity of 0.89 and a companion mass of $1.22_{-0.05}^{+0.06} \rm M_\odot$; \citealt{Ridolfi2019}), which are quite unlike the orbital eccentricities and companions to MSP binaries observed in the Galactic disk. 

As essential targets for searching pulsars, a number of GCs have been continuously monitored by the
Five-hundred-meter Aperture Spherical Radio Telescope (FAST; \citealt{Nan2011,Jiang2019,Jiang2020,Pan2020,Wang2020}).
The FAST GC pulsar survey was started in 2018 and has reported 43 new pulsars in 12 GCs
\footnote{\url{https://fast.bao.ac.cn/cms/article/65/}}.
Most new discoveries are binary pulsars (33 out of 43), except in NGC6517 (2 out of 13; only NGC 6517J, and possibly 6517N are binaries).

NGC 5024 (M53, $\alpha = 13^{\mbox{\scriptsize h}} 12^{\mbox{\scriptsize m}} 55.3^{\mbox{\scriptsize s}}$,
$\delta = +18\degr 10\arcmin 05\arcsec$) 
is located in the intermediate Galactic halo at a Heliocentric distance of R$_{\rm Sun} = 17.9$ kpc and
a galactocentric distance of R$_{\rm gc}$=18.4 kpc \citep{Harris1996,Harris2010}, it is the most distant GC with known pulsars. Also, with a central density of $\rho_c \sim 1.2\times 10^3 ~\rm L_{\odot} pc^{-3}$ it is the least dense of all GCs with pulsars. Furthermore, with [Fe/H]~$\sim -2.10$, it is among the most metal-poor GCs in the Galaxy \citep{Harris1996,Harris2010}.

Before this study, four pulsars (including 3 MSPs) have been reported, with a dispersion measure (DM) of $\sim$ 25 $\rm pc~cm^{-3}$. The first binary pulsar in M53, B1310+18A (M53A), was discovered using the Arecibo 305-m radio telescope, on a band 8 MHz wide centered on 430 MHz \citep{Kulkarni1991}.  Its spin period ($P$) and DM value are $\sim$ 33 ms and 24.0 $\rm pc~cm^{-3}$, respectively; it is a binary pulsar in a 256-day orbit with a low-mass
{\bf ($0.35 \, \rm M_{\odot}$) companion. Until now, no phase-coherent timing solution has been published for this pulsar.}
More recently, three new MSPs (PSRs J1312+1810B, C and D, henceforth M53B to D) were discovered by FAST, all significantly fainter than M53A \citep{Pan2021}, but no timing solutions were published for these pulsars. 
These discoveries show that, with the high sensitivity of FAST, we can discover significant numbers of pulsars, even in very distant GCs like M53. These discoveries will help us better characterize the distribution of MSPs in globular clusters, especially for those with very low metallicities.

In section~\ref{sec:observations}, we describe our 2 years of FAST observations and how the resulting data were reduced. In section~\ref{sec:results}, we present the results of our observations - the discovery of M53E and the determination of timing solutions for all pulsars - and discuss their implications.
We summarize our findings in section~\ref{sec:conclusions}.

{\bf Using FAST data, we have also derived a phase-coherent timing solution for M53A, and have extended it to the early 1990's using archival Arecibo data, obtaining very precise spin, astrometric and orbital parameters. However, some archival data has not yet been fully analyzed. The results of this work will be published by Lian et al. (in prep).}


\begin{table*}[ht]
	\centering
	\caption{Observation details. }
	\label{GCs}
	\begin{scriptsize}
	\setlength{\tabcolsep}{4mm}{
		\begin{tabular}{ccccc}
			\hline
			Date         &  MJD    &  Observation length  &   M53D Detected$^{1}$    &  Sensitivity  \\
			(yyyy-mm-dd) &         &      (min)           &     Y/N                  &  $\mu$Jy      \\
			\hline
			2019 Sep 30   &  58817     &  300  &  Y   &  0.48 \\
			2020 Oct 26   &  59148     &  300  &  Y   &  0.48 \\
   		2020 Nov 19   &  59172     &  300  &  Y   &  0.48 \\
			2020 Nov 20   &  59173     &  300  &  Y   &  0.48 \\
			2020 Nov 21   &  59174     &  300  &  Y   &  0.48 \\
			2020 Nov 23   &  59176     &  300  &  Y   &  0.48 \\
			2020 Nov 25   &  59178     &  300  &  Y   &  0.48 \\
			2020 Dec 04   &  59187     &  300  &  N   &  0.48 \\
			2020 Dec 28   &  59211     &  300  &  N   &  0.48 \\
			2021 Jan 08   &  59222     &  240  &  N   &  0.54 \\
			2021 Feb 12   &  59257     &  300  &  Y   &  0.48 \\
			2021 Feb 13   &  59258     &  300  &  Y   &  0.48 \\
			2021 Jun 05   &  59370     &  288  &  N   &  0.49 \\
			2021 Jun 11   &  59376     &  180  &  N   &  0.62 \\
			2021 Sep 20   &  59477     &  300  &  Y   &  0.48 \\
			2022 Feb 24   &  59634     &  180  &  Y   &  0.62 \\
			2022 Mar 06   &  59644     &  180  &  N   &  0.62 \\
			2022 Mar 08   &  59646     &  180  &  N   &  0.62 \\
			2022 Mar 23   &  59661     &  150  &  Y   &  0.68 \\
			2022 Mar 24   &  59662     &  186  &  Y   &  0.61 \\
			2022 Mar 28   &  59666     &  120  &  N   &  0.76 \\
			2022 Apr 05   &  59674     &  58   &  Y   &  1.09 \\
			\hline
	\end{tabular}}
    \end{scriptsize}
	\begin{itemize}
	    \item[1] M53A to C and E can be found in every observation. 
                 For M53D detection, its signal can be seen by folding data with timing solution. 
	\end{itemize}
\end{table*}

\begin{figure*}
\centering
	\includegraphics[width=1.0\textwidth,height=0.22\textwidth]{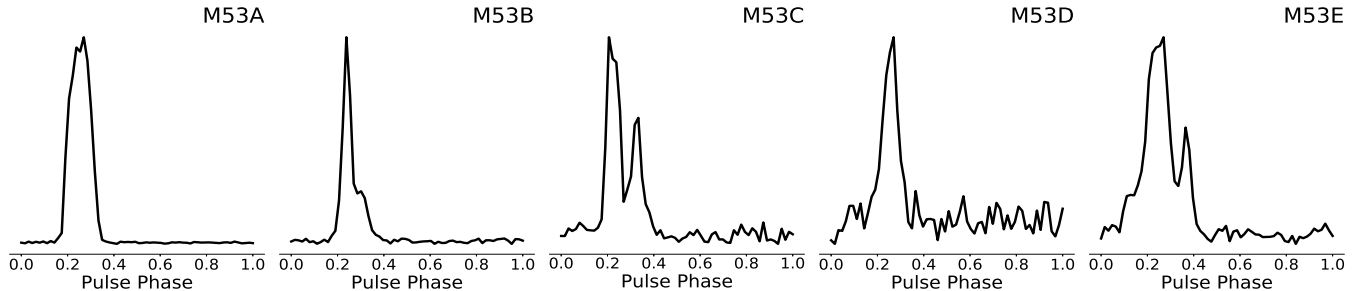}
	\caption{Integrated pulse profiles by summing all detections. 
             All profiles are for 64 bins.}
\end{figure*}

\section{Observation and Data reduction}
\label{sec:observations}

The tracking observation of M53 was initiated on November 30$^{th}$ 2019, as the pilot survey for the FAST GC pulsar survey and $\rm SP^{2}$ project 
\footnote{Search of Pulsars in Special Population, \url{https://crafts.bao.ac.cn/pulsar/SP2/}} \citep{Pan2021}.

For this observation, we used the center beam of the FAST 19-beam L-band receiver, which has a beam size of $\rm \sim 3 ^\prime$ and covers a frequency range of 1.0-1.5 GHz. The data were 8-bit sampled for two polarizations and channelized into 4096 channels (0.122 MHz channel width), the resulting power spectra were summed every 49.152 $\mu \rm s$. The pulsars, M53B to D were discovered in this observation, with dispersion measures (DMs) of  26.0, 26.2, 24.6, and 25.9 $\rm pc~cm^{-3}$ respectively. The timing observations were then carried out until 2022, with the same receiver and back-end setup as for the discovery observation, which is suitable for the analysis of multiple pulsars and allows additional discoveries; their details are presented in Table~\ref{GCs}.

This table lists the search sensitivity for these observations. This was calculated using the radiometer equation \citep{Lorimer2004}:
\begin{equation}
S_{\rm min} = \beta \displaystyle  \frac{(S/N_{\rm min}) T_{\rm sys}}{G \sqrt{n_{\rm p} T_{\rm obs} \Delta f}}\sqrt{\frac{W}{P-W}},
\end{equation}
where $\beta$ is the sampling efficiency and equals 1 for our 8-bit recording system, the minimum signal-to-noise (S/N) ratio ($S/N)_{\rm min}$ is 10, the system temperature ($T_{\rm sys}$) is 24~K, the antenna gain ($G$) is 16~K~Jy$^{-1}$, the number of polarizations ($n_{\rm p}$) is 2.  T$_{\rm obs}$ is the integration time (in seconds) and $\Delta f$ is the bandwidth in MHz, for this survey it is 300~MHz. 
The $W$ and $P$ are pulse width and period, respectively, and we take 10\% for $W/P$.

In order to search for additional pulsars, the search on timing observation data was done on all the data with the PulsaR Exploration and Search TOolkit \citep[{\sc presto}, ][]{Ransom2001,Ransom2002,Ransom2003}).
The acceleration search with {\sc presto}'s $accelsearch$ routine was first without acceleration ($z_{\rm max}=0$) and then  with a middle acceleration ($z_{\rm max}=200$). The parameter $z_{\rm max}$ represents the maximum number of Fourier bins that the signal can drift (linearly with time) in the power spectra (e.g., due to orbital motion or pulsar spin down over the course of the observation, see e.g., \citealt{Ransom2002}). 

To search for short-orbit pulsars, the data were also separated into 1-hr segments, which were then searched separately. This resulted in the discovery of M53E, which is a binary pulsar in a 2.4-day orbit. The signal of this pulsar was later recovered in the 2019 data. Because the observation time is $\rm \sim$10\% of the orbital period, it was missed in the previous search. Its discovery indicates that the 10\% orbital period can be the limitation for acceleration searches, as already extensively discussed in the literature. To find short orbital period binaries, the search should either be done with short observations or new methods (e.g., jerk search \citep{Andersen2018}) should be employed.

\begin{table*}
	\centering
	\caption{Timing solutions for previously discovered pulsars M53B to D and new pulsar M53E.
		For M53D and E, the eccentricity is derived from the ELL1 parameters$^a$.}
	\label{table:timing_known}
        \setlength{\belowdisplayskip}{3pt}
	\begin{footnotesize}    
		 \resizebox{\textwidth}{55mm}{
			\begin{tabular}{lcccc}
				\hline\hline
				Pulsar name      &      M53B       &      M53C      &      M53D      &      M53E      \\
				\hline
				MJD range          & 58817---59674   & 58817---59674  &  58817---59674  & 58817---59674 \\
    			Reference epoch (MJD)       &   59477.1528     &  59477.1528  & 59477.1528      &   59477.1528   \\
				Number of TOAs   &     344         &     250    &     66         &     319        \\
                EFAC         &     1.153        &     1.385    &     1.065        &     1.028  \\
				Timing residual r.m.s. ($\mu$s)  &     19.24        &     88.57   &     53.23      &     52.98        \\
                Reduced $\chi^2$        &     1        &     1    &     1        &     1  \\
				Solar System ephemeris model    &  DE440   &  DE440   & DE440    &  DE440  \\
				Binary model      &   DD     &  ---     & ELL1    &  ELL1     \\    
				\hline
				\multicolumn{4}{c}{Measured quantities} \\
				\hline

				Right Ascension, $\alpha$ (J2000)   &  13:12:54.7219(1)      &  13:12:55.6624(3) &  13:12:55.6338(6)   &  13:12:55.8311(4)       \\
				Declination, $\delta$ (J2000)       &  +18:10:03.231(3)      &  +18:10:09.549(9)   &  +18:09:55.05(2)    &  +18:10:17.233(7)    \\

                Pulse Frequency, $f$ (Hz)        &  160.19831544487(2)  &  79.77612072867(4)    & 164.7517730112(2)       &  251.75352627488(9) \\
				Pulse Frequency Derivative, $\dot{f}$ (s$^{-2}$)    &  $4.48(1) \times 10^{-16}$    &  $-3.35(2) \times 10^{-16}$  & $-6.14(7) \times 10^{-16}$   &  $4.80(5) \times 10^{-16}$   \\
    			Dispersion measure, DM (cm$^{-3}$~pc)   &   25.960(1)       &    26.194(6)   &   24.61(1)            &   25.879(5)   \\

				Orbital Period, $P_{\rm b}$ (day)             &  47.6773506(1)      &  ---     & 5.75023947(8)     &  2.43137859(1)    \\
				Projected Semi-major Axis, $x$ (lt-s)   &  22.616556(2)    &  ---     & 5.91397(1)   &  2.371683(5)  \\
				Orbital eccentricity, $e$        &   0.0132417(2)      &  ---   & $1.4(4) \times 10^{-5}$$^a$  &   $0.9(5) \times 10^{-5}$$^a$     \\
				Longitude of periastron, $\omega$ (deg) &   26.8885(6)        &  ---  & ---   &   ---     \\
 				Epoch of passage at Periastron, $T_0$ (MJD)  &  59653.08363(8)   &  ---  & ---     &  ---   \\
				Epoch of passage at Ascending Node, $T_\textrm{asc}$ (MJD)   &    ---    &    ---  & 59259.173859(3)  &  59643.937510(2)  \\
				EPS1  & --- & ---  &  $1.2(4) \times 10^{-5}$ & $0.8(5) \times 10^{-5}$ \\
				EPS2  & --- & --- & $0.6(5) \times 10^{-5}$ & $0.5(5) \times 10^{-5}$ \\     

				\hline
				\multicolumn{4}{c}{Derived quantities} \\
				\hline
    			Spin Period, $P$ (ms)    &   6.2422628928589(7)  &    12.535079305264(6)   &    6.069737409941(8)      &    3.972138999587(1)   \\
				1st Spin Period Derivative, $\dot{P}$ (s s$^{-1}$)    &  $-1.746(4)\times 10^{-20}$    &  $5.26(3)\times 10^{-20}$   &  $2.26(3)\times 10^{-20}$    &  $-7.58(8)\times 10^{-21}$  \\
                    Mass Function, $f_{({M_p})}$ (${\rm M}_\odot$)    &   0.005464887(1)    &   ---  &   0.00671726(5)    &   0.00242321(1)     \\  
				Minimum companion mass, $M_{\rm c, min}$ (${\rm M}_\odot$)     &   0.2455   &   ---   &   0.2650   &   0.1824    \\  
                    Angular offset from centre in $\alpha$, $\theta_{\alpha}$ (arcmin)  &  $-$0.1261&   +0.0974&  +0.0913&   +0.1374\\  
                    Angular offset from centre in $\delta$, $\theta_{\delta}$ (arcmin)  &  $-$0.0363&   +0.0691&  $-$0.17238&   +0.1971\\ 
                    Total angular offset from centre, $\theta_{\perp}$ (arcmin)  &  0.1312&   0.1195&  0.1951&   0.2403\\ 
                    Total angular offset from centre, $\theta_{\perp}$ (core radii)  &  0.3750&   0.3413&  0.5573&   0.6865\\ 
                    Projected distance from centre, $r_{\perp}$ (pc)  &  0.7063&   0.6428&  1.0497&   1.2929\\ 

				\hline
		\end{tabular}}
	\end{footnotesize}
\end{table*}

We used the fitorb.py routine \footnote{\url{https://github.com/scottransom/presto/blob/master/bin/fitorb.py}} to obtain the initial orbital parameters for the binary pulsars. 
From our detections, we derived pulse times of arrival (ToAs) by cross-correlating the pulse profiles against a high-S/N template, which was obtained from fitting a set of Gaussian curves to the best detections.
For the subsequent analysis of the ToAs, the {\sc tempo} pulsar timing package \citep{Nice2015} \footnote{\url{http://tempo.sourceforge.net}} was carried out. With the latter program, we used the {\sc dracula} script\footnote{\url{https://github.com/pfreire163/Dracula}} \citep{Freire2018} to derive phase-connected timing solutions for all the pulsars. As seen from Table \ref{GCs}, the gap between some observations can be one month or even much longer; this sparsity made it impossible to determine the timing solutions of pulsars M53B to E without recourse to {\sc dracula}. 


\section{Results and Discussion}
\label{sec:results}

\begin{figure*}
	\begin{center}
		\includegraphics[width=1\textwidth,height=0.6\textwidth]{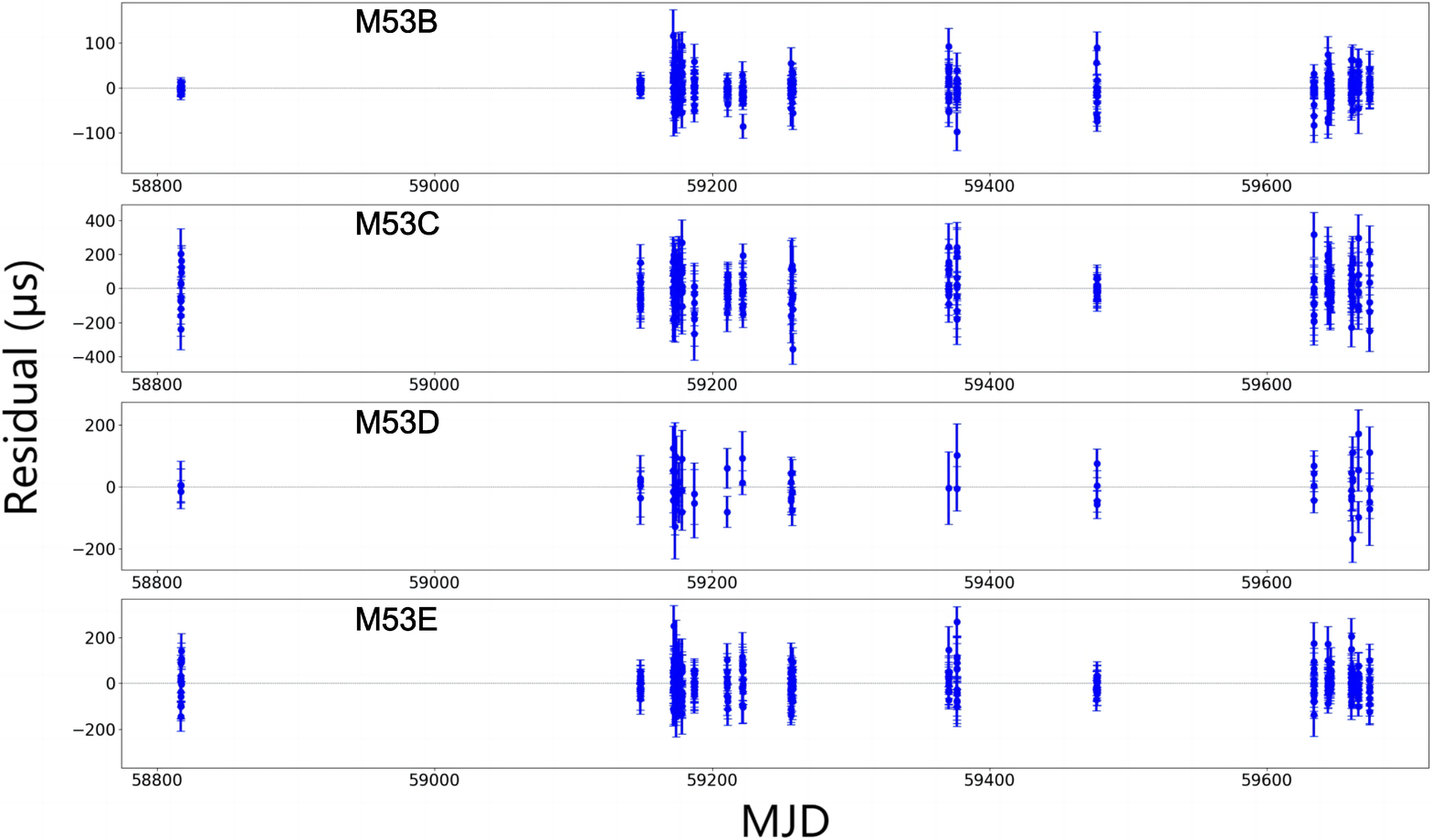}
	\end{center}
	\caption{Timing residuals from the best-fit timing models presented in Table 2 as a function of the observation date for M53B to E.}
        \label{fig2}
\end{figure*}

In Table~\ref{table:timing_known}, we present the timing solutions for M53B to E. In Fig.~\ref{fig2}, {\bf we show the post-fit timing residuals with all of the TOAs, obtained by these solutions}. The fact that these show no trends indicates that the solutions provide a good description of the ToAs.

\subsection{Properties of pulsars in M53}

{\bf M53A is a relatively slow 33 ms pulsar, in a rather wide 256-day orbit.}
All new pulsars are significantly fainter than the previously known pulsar M53A, {\bf spin faster and, for those in binaries, have shorter orbital periods}. M53B, the second brightest pulsar known in this cluster, is an MSP with $P = 6.24$ ms in a 47.7-day, mildly eccentric ($e = 0.0132$) binary system with a low-mass companion: assuming a pulsar mass of $1.4 ~ \rm M_{\odot}$, the minimum companion mass ($M_{\rm c, min}$) is $0.246 ~ \rm M_{\odot}$.

M53C is much fainter than M53B. It is an isolated MSP with $P = 12.53\,$ ms, which helped detecting it in our observations.
M53D is the faintest pulsar with possible interstellar scintillation. It is an MSP with $P = 6.07$ ms and a low-eccentricity orbit ($e \sim 0.000014(4)$) with a period of 5.75 d, again with a low-mass companion (with the assumptions above, $M_{\rm c, min} = 0.265 \rm M_{\odot}$). With FAST, we can detect M53B, C, and E in each observation; the detection rate of M53D is $\sim$ 64\%. Finally, our newly discovered pulsar, M53E, is a 3.97-ms pulsar in a 2.43-day, low-eccentricity orbit, again with a low-mass companion $M_{\rm c, min} = 0.182 ~ \rm M_{\odot}$.


\begin{figure*}
\centering
	\includegraphics[width=\columnwidth]{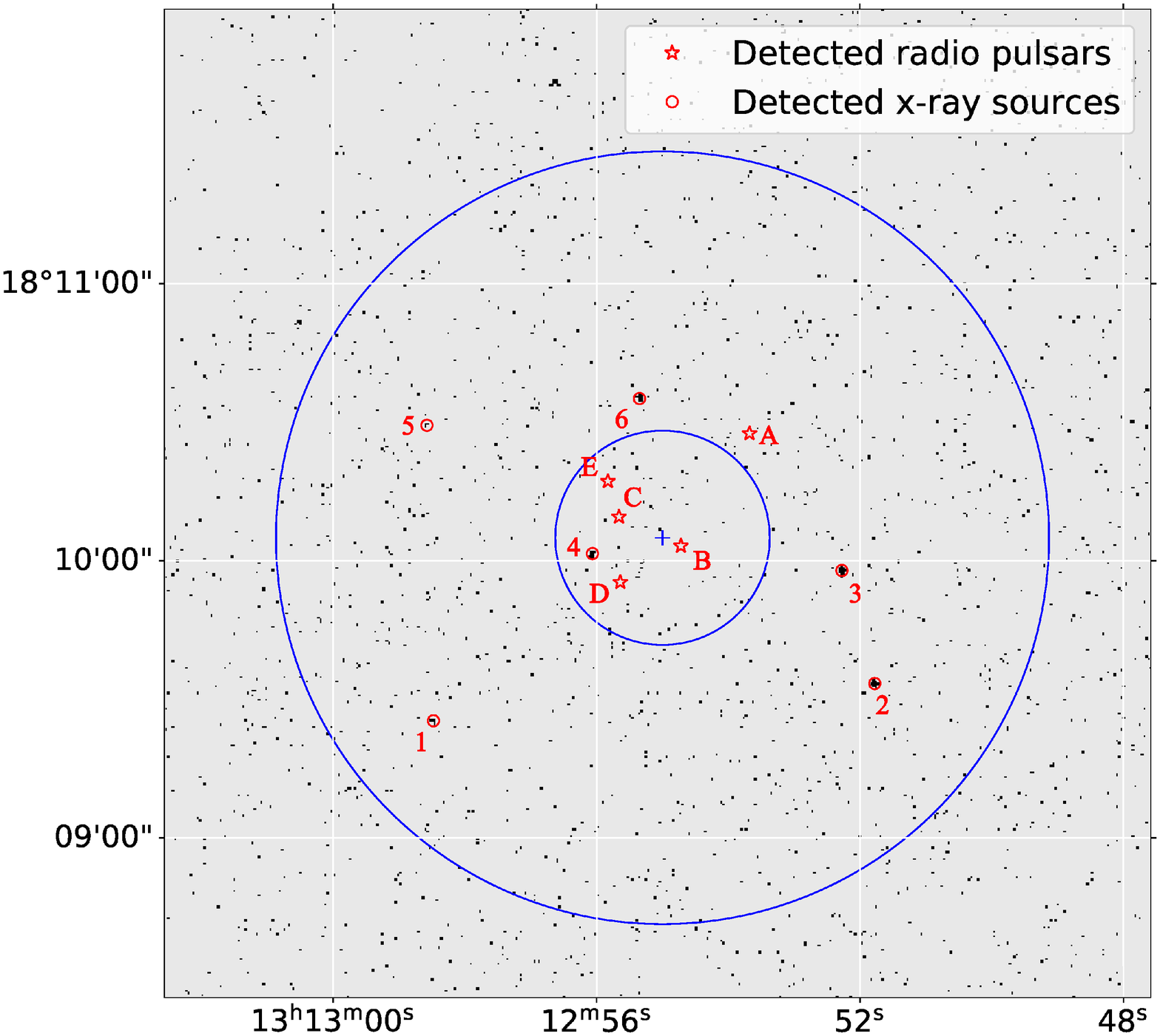} \includegraphics[width=\columnwidth]{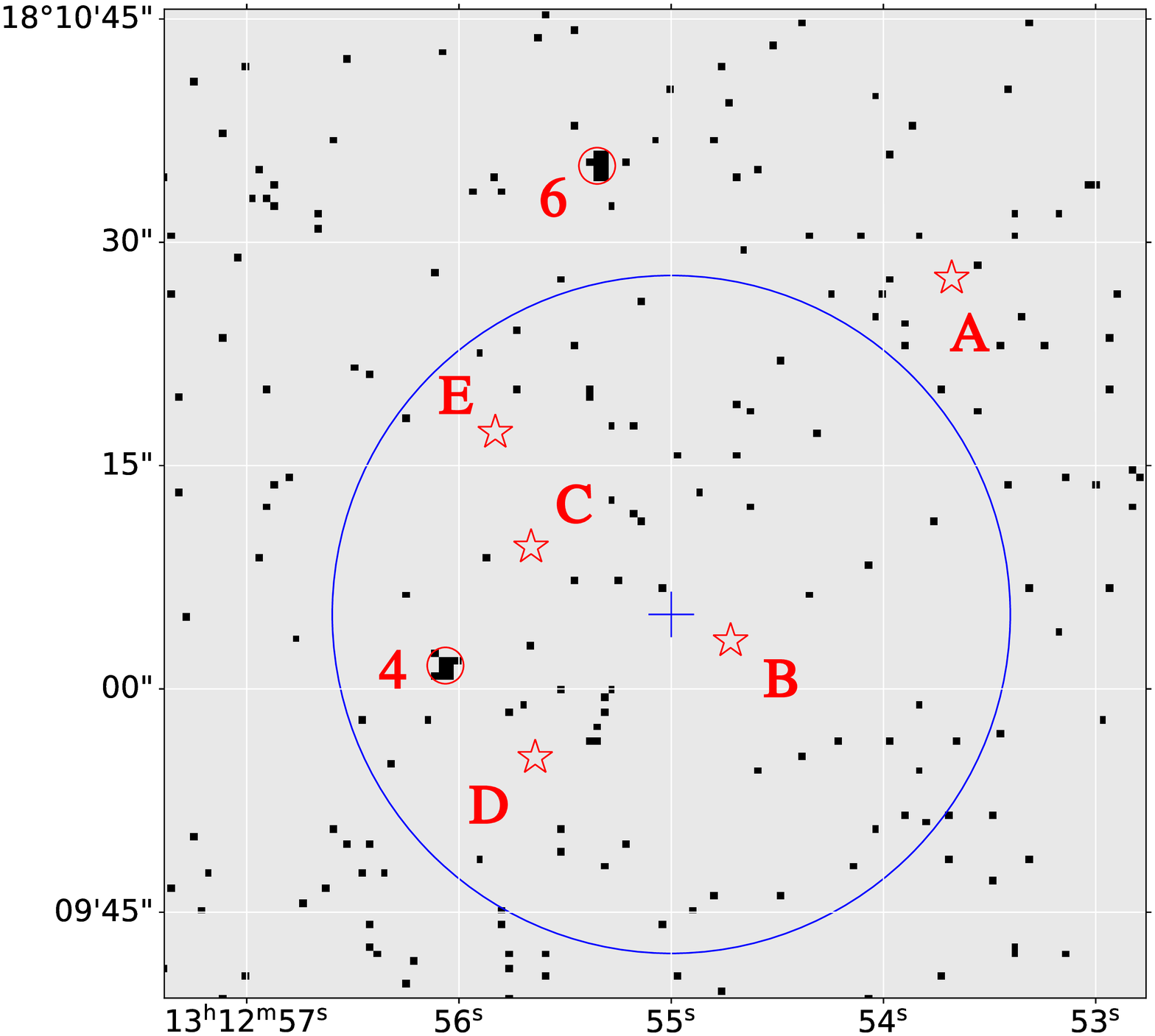}
	\caption{Positions of M53A to E, and six X-ray sources in M53, marked with red stars and circles respectively. The center of M53 is shown as a blue cross. {\bf The position of M53A is from Lian et al. (in prep.).} The blue circles in the middle show a core radius (0.35\arcmin) and a half-light radius (1.31\arcmin). The background is the X-ray image of M53 from the $Chandra ~ X-ray ~ Observatory$ archive (OBsID 6560). The left panel displays the details. For reference, the beam size of FAST is 3\arcmin.}
        \label{fig3}
\end{figure*}

Fig.~\ref{fig3} displays the positions of \textbf{M53A to E} in the cluster. 
All the new discoveries are located within 0.24\arcmin\, from the centre of M53, well within the core; in principle, the beam size of 3\arcmin\, should allow discoveries significantly farther out, at least to 1.5\arcmin, which is already beyond the half-light radius.
Thus, the strong concentration of new discoveries in the core is not an effect of the small beam size.
The positions of six bright X-ray sources in M53 detected by $Chandra ~ X-ray ~ Observatory$ (CXO) are presented as red circles \citep{Zhao2022}. Only one source (number 4) is located within the core region. With our best-fitted positions with 1$\sigma$ uncertainty($\sim$ 10$^{-2}$ arcsec for RA and Dec),  no relevant X-ray counterparts of M53B to E have been found. An unabsorbed X-ray luminosity for the six sources is within the range $3.3 \times 10^{31}$ to $7.6 \times 10^{32}$ $\rm erg \; s^{-1}$ in the band 0.3–8 keV, which lies above the estimated limiting luminosity of $3 \times 10^{31}$ $\rm erg \; s^{-1}$ \citep{Zhao2022}. 

\textbf{The X-ray emission of binary MSPs is observed to be both thermal and nonthermal character. For example, the X-ray radiation of most MSPs (at least 20 out of 29 MSPs) in 47 Tuc is well-defined by a thermal model, which is believed to originate from the magnetic polar caps of the underlying neutron star \citep{Bogdanov2006,Bhattacharya2017}. Alternatively, the nonthermal X-ray of B1957+20 and three radio eclipsing binaries MSPs 47 Tuc J, O, and W indicate the shock where the winds of the pulsar and its companion collide \citep{Stappers2003,Bogdanov2006}. Furthermore, in some more extreme cases, transient X-ray emission is produced by the accretion of matter from a low-mass companion star (e.g., see XTE J1808$-$369, \citealt{Wijnands1998}). None of the binaries in M53 shows eclipses, so we’d generally expect only the thermal X-ray emission as in most pulsars in 47~Tuc. The non-detection of the M53 pulsars in the band 0.3–8 keV shows that their thermal X-ray emission is, when seen at the large distance of M53, too faint to be detectable in existing observations.}

\subsection{The single-binary encounter rate}

The stellar encounter rate $\Gamma \propto {\rho_c}^2{r_c}^3/v$ \citep{Verbunt2003} is $\Gamma_{\rm M4}=0.84$ (where $\rho_c$ is the central density of the GC, $r_c$ is cluster core radius, $v$ is the velocity dispersion, the result is normalized to the values of the GC M4, as in \citealt{Verbunt2014}). The single-binary encounter rate ($\gamma_{\rm M4} = \propto \rho_c/v$) is $0.21$. $\Gamma$ ranks as the fourth lowest, and $\gamma$ is the second lowest among the 38 GCs with known pulsars. M53 has four binary pulsars and one isolated pulsar, in this it is similar to the other GCs with a low $\gamma_{\rm M4}$, which also have a high fraction of binary pulsars like M3 ($\gamma_{\rm M4}=0.6$, five binary pulsars),  M13 ($\gamma_{\rm M4}=0.5$, four binaries and two isolated pulsars), 
and M71 ($\gamma_{\rm M4}=0.4$, five binary pulsars).  This is consistent with the theoretical expectation that in clusters with a small $\gamma_{\rm M4}$ X-ray binaries are expected to evolve without much disturbance, forming binary MSPs similar to those of the Galactic disk. As noticed by \cite{Verbunt2014}, and confirmed several times since (e.g., in NGC 6624, \citealt{Abbate2022} and the many isolated pulsars found in NGC~6517 by FAST mentioned in the Introduction), for the GCs with the largest values of $\gamma_{\rm M4}$ - the core-collapsed clusters - the situation is very different, with the population of detectable pulsars being dominated by isolated objects, a consequence of the high rates of orbital disruption.

\subsection{Eccentricities of the M53 binary pulsars}

During the formation of MSP-WD binaries, any eccentricity that the system might have prior to the low-mass X-ray binary phase will dissipate quickly from tidal circularisation, as observed in Galactic MSP binaries, for which the eccentricities follow the relation  predicted by \cite{Phinney1992}. This means that the much larger orbital eccentricities observed for binary MSPs in globular clusters are likely due to gravitational perturbations of passing stars after accretion ceases \citep{Rasio1995,Heggie1996}.

In denser GCs, there will be many more encounters per unit time: for the example above, Terzan 5, the single binary encounter
rate is $\gamma_{\rm M4} = 13.0$, this means that within any time interval there is a $\sim 62$ times higher probability of a
close encounter Terzan 5 compared to M53. 
This explains why the binaries in Terzan 5 have such high eccentricities \citep{Prager2017,Martsen2022} compared to M53 where the orbits of the previous binary system (M53A) and the new binary pulsars are all mildly eccentric, with that of M53B having the largest eccentricity ($e=0.013$).

For orbital periods of a few days to a few hundred days, \cite{Phinney1992} predict eccentricities ranging from $\sim 10^{-6}$ to $\sim 10^{-3}$. The eccentricities of the binary systems in M53 are still significantly higher than this.
Given the low stellar density of the cluster, it is important to verify whether the eccentricities of its binaries can arise from interactions with the stellar population of the cluster, which we do now. The time scale required to generate the eccentricity of M53B can be evaluated as \citep{Rasio1995,Lynch2011}:
\begin{eqnarray}
\nonumber t_{>e} &&\simeq 4 \times 10^{11}\; \yr 
\left (\frac{n}{10^4\; \pc^{-3}} \right )^{-1} 
\left (\frac{v}{10\; \km\, \ps} \right ) \\
&&\times\left (\frac{P\rmsub{b}}{\mathrm{days}} \right)^{-2/3}
e^{2/5},
\end{eqnarray}
where $n$ is the number density of stars ($n \propto \rho_c$, $\rho_c$ is the density of GC), 
$v$ is the one-dimensional core velocity dispersion ($v = 4.4 \; \rm km s^{-1}$ for M53; \citealt{Harris2010}), 
and $P\rmsub{b}$ is the orbital period.
Normalized with the values $\rho_c \sim 1.95\times 10^5 \; \rm L_{\odot} \, \rm pc^{-3}$ and $n \approx 4.7 \times 10^5 \; \rm pc^{-3}$ of NGC 6517 \citep{Lynch2011}, the number density $n$ is roughly estimated as $n \approx 2.6 \times 10^3 \, \rm  pc^{-3}$,
through the luminosity density of M53 ($\rho_c \sim 1.17\times 10^3 \; \rm L_{\odot} \, \rm pc^{-3}$) reported in \citet{Harris2010}.
These values imply $t_{>e} \approx 9.1 \; \times \; 10^9$ years, which is within the age of M53 $\sim$ 12.67 Gyr \citep{Forbes2010}. We note that the lower limit for the characteristic age of M53B, discussed below, is $\sim 20 \, \rm Gyr$, a further indication that the pulsar is very old.

The eccentricities of M53D and M53E are fitted using the binary model ELL1 \citep{Lange2001}, both are below $10^{-4}$. Although they are one order of magnitude larger than the prediction of \cite{Phinney1992}, they are smaller than expected for their orbital period and the density of their local environment. This could possibly be due to them either being younger than M53B or their orbits carrying them through less dense regions of M53.

\subsection{Pulsar acceleration caused by cluster dynamics}

\begin{figure}
	\includegraphics[width=\columnwidth]{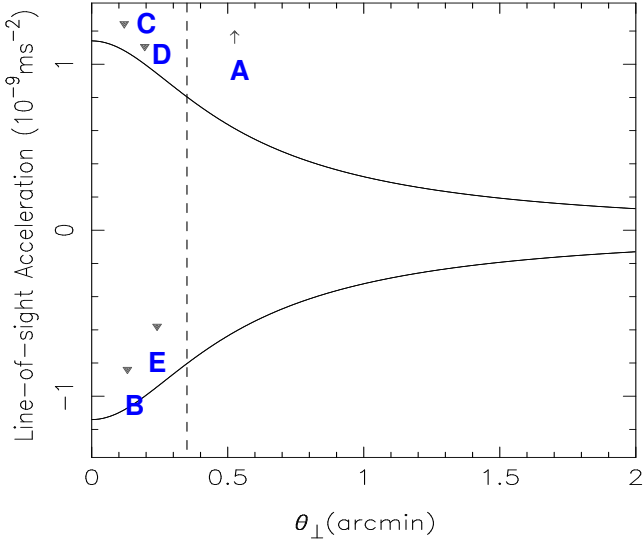}
	\caption{Acceleration model for M53. The black solid lines represent the upper and lower limits for the line-of-sight accelerations ($a_{\ell \rm GC}$) caused by the cluster as a function of the total angular offset from the centre of the cluster ($\theta_{\perp}$). The triangles pointing down represent independent upper limits for the pulsar accelerations along the line of sight, $a_{\ell, \rm P, max}$.
 {\bf The limit for M53A, from Lian et al. (in prep.), is well above the upper limit of this Figure.} The vertical dashed line represents the core radius. The cluster model can account for the negative $\dot{P}$ values of M53B and M53E. The remaining pulsars are above that prediction, which is allowed since $a_u$ has a contribution from their instrinsic spin period derivatives, which are positive.}
        \label{fig4}
\end{figure}

In Table~\ref{table:timing_known}, we see that two of our discoveries (M53B and E) have negative $\dot{P}$. Since the intrinsic spin period derivative $\dot{P}_{\rm int}$ is positive for radio pulsars (which are generally spin powered), an observation of a negative $\dot{P}$ suggests that the pulsar is accelerating in a gravitational field. The observed spin period derivative is generally given by:
\begin{equation}
\left( \frac{\dot{P}}{P} \right)_{\rm obs} = \left( \frac{\dot{P}}{P} \right)_{\rm int} + \frac{\mu^2d}{c} + \frac{a_{\ell,\, \rm GC}}{c} + \frac{a}{c},
\end{equation}
where $\mu$ is the total proper motion of the system, $d$ is the distance to the cluster ($\mu^2d / c$ is so-called the Shklovskii effect; see \citealt{Shklovskii1970}), $c$ is the speed of light, $a_{\ell,\, \rm GC}$ is the line-of-sight acceleration of the pulsar in the gravitational field of the cluster, and \textbf{$a$ is the acceleration of the center of mass of the globular cluster in the potential of the Galaxy minus the Galactic acceleration of the Solar system projected along the line of sight from the Earth to M53.}

Of these effects, two can be estimated with some precision: using the galactic rotation model developed by \cite{Prager2017}, we obtain
$a \, = \, 2.999 \times 10^{-12}\, \rm m \, s^{-2}$. The proper motions of the pulsars have not been measured yet, however, they should be very similar to that of M53, $0.042 \pm 0.08 ~\rm mas ~ yr^{-1}$ (from $Gaia$ DR2 data, \citealt{Baumgardt2018}); thus $\mu^2d$ is $2.303 \times 10^{-14} \rm m \, s^{-2}$. Compared with the observed pulsar accelerations in M53 (e.g., $c \dot{P}_{\rm obs}/P > -5.789 \times 10^{-10} \rm ~ m ~ s^{-2}$ for M53E), these effects are negligible.

\begin{table*}[!htpb]
    \centering
    \setlength{\tabcolsep}{4mm}{
        \begin{tabular}{cccccc} 
            \hline
            Pulsar name & $a_{\ell, \rm P, max}$ & $a_{\ell, \rm max}$ & $\dot{P}_{\rm int}$ & $B$ & $\tau_{\rm c}$ \\
                & $ 10^{-9} \rm m \, s^{-2}$ & $ 10^{-9}\, \rm  m \, s^{-2}$ &  $ 10^{-20} \, \rm s \, s^{-1}$ & $10^9$ G & Gyr  \\
            \hline
            M53B & $-$0.84 & 1.07 &  $\textless$0.48  & $\textless$0.18 &  $\textgreater$20.45 \\
            M53C & 1.24  & 1.08 & 0.68  $\sim$ 9.72   & 0.29 - 1.11  &   29 - 2.04  \\
            M53D & 1.11  & 1.00 & 0.22  $\sim$ 4.26   & 0.12 - 0.51  &  45 - 2.26  \\
            M53E & $-$0.58 & 0.94 & $\textless$0.48   & $\textless$0.14 &  $\textgreater$13.03  \\
            \hline
        \end{tabular}}
    \caption{For each of the new pulsars in M53, we calculate the upper limit for the pulsar accelerations, the upper and lower limits for the line-of-sight pulsar acceleration due to the cluster potential and the resulting limits on the intrinsic spin period derivative, the surface magnetic field strength ($B$), and the characteristic age ($\tau_c$) respectively.
    \label{accelerations}}
\end{table*}

The dominant term is the acceleration caused by the field of the GC, $a_{\ell, \rm GC}$. To model this, we used an analytical model of the cluster described by \citet{Freire2005}, which assumes the \citet{King1962} density profile. With this, the accelerations due to the cluster potential at $x$ (the distance from the pulsar to the centre of the GC divided by its core radius $r_c=\theta_c d$) can be calculated as:
\begin{equation}
a_{\rm GC}(x) = \frac{9 \sigma^2}{d \theta_c} \frac{1}{x^2}
\left( \frac{x}{\sqrt{1+x^2}}  - \sinh^{-1}x \right).
\end{equation}
The parameters for this model include the position and distance of M53 (see section~\ref{sec:introduction}), its core radius ($\theta_c \, = \, $0.35\arcmin; \citealt{Harris2010}), and the central stellar velocity dispersion ($\sigma \, \sim \, 6.5 ~\rm km ~ s^{-1}$)\footnote{\url{https://people.smp.uq.edu.au/HolgerBaumgardt/globular/}}. The values for the line-of-sight component $a_{\ell, \rm GC}(x)$ can be calculated by multiplying $a_{\rm GC}(x)$ by $\ell / x$, where $\ell$ is the distance (also in core radii) to the plane of the sky that passes through the centre of the cluster. In Fig.~\ref{fig4}, the solid black curves represent the maximum and minimum values of $a_{\ell, \rm GC}(x)$ for each line of sight through the cluster, which is characterized by a constant angular offset from the centre, $\theta_{\perp}$.

For all pulsars, we can derive an independent upper limit on the acceleration from $\dot{P}_{\rm obs}$:
\begin{equation}
a_{\ell, \rm P, max} \, = \, c \frac{\dot{P}_{\rm obs}}{P} 
\end{equation}
Fig.~\ref{fig4} graphically shows the constraints for each pulsar in M53 as the triangles pointing down, which highlights that they are upper limits on the cluster acceleration that assume $\dot{P}_{\rm int} = 0$. These also presented in Table~\ref{accelerations}.

From this figure, we conclude that, despite the small accelerations predicted by the analytical model described above, it can account for the negative $\dot{P}_{\rm obs}$ of M53B and E. It cannot fully account for the positive $ a_{\ell, \rm P, max}$ of the remaining pulsars, but this is to be expected because their spin-down has a contribution from a positive $\dot{P}_{\rm int}$. 

Taking the maximum and minimum accelerations caused by the gravitational field of the GC for the line of sight of each pulsar, we can calculate maximum (and in some cases minimum) limits for the $\dot{P}_{\rm int}$ of each pulsar; from these we can calculate maximum (and in some cases minimum) values for their magnetic fields and minimum (and in some case maximum) values for their characteristic ages. These values are presented in Table~\ref{accelerations}.

From these values, it is clear that all new pulsars in M53 have weak magnetic fields and are very old, all with characteristic 
ages larger than 2 Gyr. This makes them, again, similar to the MSP population observed in the Galactic disk, as predicted by the low $\gamma_{\rm M4}$.

\section{Conclusions}
\label{sec:conclusions}

  In this paper, we present the discovery of M53E during the monitoring observations.
  With 22 FAST observations from 2019.11 to 2022.04, we have obtained timing solutions of M53B to E, these were only possible owing to the high sensitivity of the radio telescope. All of these are millisecond pulsars; M53C is the only isolated pulsar known in M53. 
  Based on the orbital characteristics,  the orbiting companions for M53B, D, and E are most likely white dwarf stars
  with masses of 0.25, 0.27, and 0.18 ${\rm M}_\odot$, respectively. All of them have low orbital eccentricities.
  
  Two of the new pulsars have negative $\dot{P}_{\rm obs}$. Using an analytical cluster potential model for M53 (see e.g., \citealt{Freire2005}), we see that, despite its small predicted accelerations, it can account for those $\dot{P}_{\rm obs}$. We conclude, additionally, that all four pulsars have magnetic fields of the order of, or smaller than, $10^9\, \rm G$
  and characteristic ages of several Gyr.

  None of the bright X-ray sources detected with the achieved $Chandra ~ X-ray ~ Observatory$ observations correspond to the
  positions of these four MSPs, no significant new sources are detectable at their positions. 
  
  The pulsar population of M53 has the characteristics expected for a GC with a low single-binary encounter rate $\gamma_{\rm M4}$, i.e., that they originate in X-ray binaries that evolved to MSP binaries without disturbance. This population thus resembles the MSP population in the Galactic disk: 1) The fraction of single pulsars is small (one isolated pulsar for four binary pulsars), 2) low eccentricity orbits (which are nevertheless slightly perturbed by nearby encounters, especially for the wider binaries, as expected) and 3) small magnetic fields and large characteristic ages. This population constrasts with that of GCs with higher  $\gamma_{\rm M4}$, where we see many more isolated pulsars, high-eccentricity binaries and apparently young pulsars \citep{Abbate2022}, which can form there from the disruption of X-ray binaries \citep{Verbunt2014}.
  
  With the ongoing FAST observations on M53, we {\bf have obtained a unique timing solution for M53A, and have extended it to the past using Arecibo observations, we now have a 32-year timing baseline for this pulsar. Additional FAST and processing of additional Arecibo observations will further improve this timing solution (Lian et al., in prep.).}
  We will also extend the timing solutions of M53B to E into the future. This will eventually allow the measurement of proper motions
  and possibly the orbital period derivatives, which will provide independent measurements of the pulsar accelerations.
  The precise positions of the pulsars will allow optical and X-ray follow-up of these systems; 
  this will be made easier by the lower-than-usual stellar density of this globular cluster, 
  which will make the unambiguous identification of companion stars easier.

\section*{Acknowledgements}
This work is supported by the National Natural Science Foundation of China (Nos. 12041301, 12021003, 11633001, and 11920101003). This work made use of the data from FAST (Five-hundred-meter Aperture Spherical Radio Telescope). FAST is a Chinese national mega-science facility, operated by the National Astronomical Observatories, Chinese Academy of Sciences.

\end{document}